\def\/{\over}

\documentclass[preprint,aps,prd]{revtex4}
\newcommand{\bea}{\begin{eqnarray}}
\newcommand{\eea}{\end{eqnarray}}
\newcommand{\beq}{\begin{equation}}
\newcommand{\eeq}{\end{equation}}

\def\k{{\vec k}}

\def\/{\over}

\begin{document}


\title{Spontaneous excitation of an accelerated multilevel atom in
dipole coupling to the derivative of a scalar field }

\author{ Yunfeng Zhu, Hongwei Yu \footnote{Corresponding author} and Zhiying Zhu  }
\affiliation{
Department of Physics and Institute of  Physics,\\
Hunan Normal University, Changsha, Hunan 410081, China }

\begin{abstract}

We study the spontaneous excitation of an accelerated multilevel
atom in dipole coupling to the derivative of a massless quantum
scalar field and separately calculate the contributions of the
vacuum fluctuation and radiation reaction to the rate of change of
the mean atomic energy of the atom. It is found that, in contrast to
the case where a monopole like interaction between the atom and the
field is assumed, there appear extra corrections proportional to the
acceleration squared, in addition to corrections which can be viewed
as a result of an ambient thermal bath at the Unruh temperature, as
compared with the inertial case, and  the acceleration induced
correction terms show anisotropy with the contribution from
longitudinal polarization being four times that from the transverse
polarization for isotropically polarized accelerated atoms. Our
results suggest that the effect of acceleration on the rate of
change of the mean atomic energy is dependent not only on the
quantum field to which the atom is coupled, but also on the type of
the interaction even if the same quantum scalar  field is
considered.

\end{abstract}

\maketitle

\baselineskip=16pt

\section{Introduction}

There has been considerable interest in the radiative properties
of uniformly accelerated atoms recently
\cite{Dalibard82,Dalibard84,Audretsch94,Audretsch95A,Audretsch95B,
Passante98,Yu05,Yu06} and a new physical picture for the
spontaneous excitation of a uniformly accelerated atom has emerged
from works based upon the formalism to separate the contributions
of vacuum fluctuations and radiation reaction proposed by
Dalibard, Dupont-Roc and Cohen-Tannoudji(DDC)
\cite{Dalibard82,Dalibard84} and subsequently generalized by
others. In particular, using DDC's formalism which demands a
symmetric operator ordering of atom and field variables, one can
show that, for ground-state atoms, the contributions of vacuum
fluctuations and radiation reaction to the rate of change of the
mean excitation energy cancel exactly and this cancellation
forbids any transitions from the ground state and thus ensures
atom's stability in vacuum; while for any initial excited state,
the rate of change of atomic energy acquires equal contributions
from vacuum fluctuations and from radiation reaction and two
contributions add up to the well-known spontaneous emission rate
\cite{Dalibard82}. On the other hand, by generalizing the
formalism of DDC \cite{Dalibard82,Dalibard84} to evaluate vacuum
fluctuations and radiation reaction contributions to the
spontaneous excitation rate of an accelerated two-level atom
interacting with a scalar field in a linear monopole like
coupling, Audretsch, M\"ueller and Holzmann
\cite{Audretsch94,Audretsch95B} have shown that, if the atom is
accelerated, then the delicate balance between vacuum fluctuations
and radiation reaction is altered since the contribution of vacuum
fluctuations to the rate of change of the mean excitation energy
is modified while that of the radiation reaction remains the same.
Thus transitions to excited states for ground-state atoms become
possible even in vacuum. This result not only is consistent with
the Unruh effect \cite{Unruh}, which is closely related to the
Hawking radiation of black holes, but also provides a physically
appealing interpretation of it. Let us note here that recently a
non-perturbative approach has been adopted to study the
interaction of a uniformly accelerated detector, modelled by a
harmonic oscillator which may be regarded as a simple version of
an atom, with a quantum field in (3+1) dimensional spacetime and
some interesting new insights have been gained~\cite{Hu}.

However, for a polarized neutral atom with an electric dipole
moment, besides the monopole like interaction considered in
Ref.~\cite{Audretsch94}, the following manifestly invariant
interaction Hamiltonian, in which the atomic dipole moment is
coupled to the derivative of a scalar field, can also be
introduced
 \begin{equation}
H_I(\tau)=m^{\mu}(\tau)\,\partial_{\mu}\phi
(x(\tau))=-e\,r^{\mu}(\tau)\,\partial_{\mu}\phi
(x(\tau))\;,\label{HI}
 \end{equation}
where $e$ is the electron electric charge, $m^{\mu}(\tau)=-e
r^{\mu}(\tau)$, the atomic electric dipole moment,
$x(\tau)\leftrightarrow(t(\tau),\textbf{x}(\tau))$, the space-time
coordinates of the atom and  $r^{\mu}(\tau)$, a four vector such
that its temporal component in the frame of the atom (proper
reference frame) vanishes and its spatial components  are given by
$\textbf{r}(\tau)$. This choice of $r^{\mu}(\tau)$ is based upon
the atom's stationarity in the proper reference frame and it
ensures that if each component of the dipole moment is meat to
evolve in time in the same way as the Unruh-DeWitt's monopole,
then  the direction of the dipole moment is kept fixed with
respect to this frame and no extra time dependence will be brought
in by the rotation of the dipole moment besides the intrinsic time
evolution~\cite{Takagi86}.

One may wonder happens  to the radiation properties of accelerated
atoms found in Ref.~\cite{Audretsch94} when the monopole like
coupling is replaced by the above dipole derivative  coupling.
This is what we are going to examine in the present paper. We will
separately calculate the contributions of vacuum fluctuations and
radiation reaction to the rate of variation of the atomic energy
of a multilevel atom in dipole coupling to the derivative of a
real massless scalar field.  Let us note here that the derivative
coupling of the form (\ref{HI}) was first considered for the
dipole detector as an alternative to the Unruh-DeWitt monopole one
\cite{Hinton83}.

\section{Interaction between a multilevel atom and the derivative of a scalar field}

So, we will consider, in this paper, a multilevel atom interacting
with the derivative of a scalar field. The Hamiltonian that
controls the time evolution of the atom with respect to the proper
time $\tau$ is written as
 \beq H_A (\tau) = \sum_{n}
\omega_{n}\sigma_{nn}(\tau)\;, \label{HA}
 \eeq
 where $|n\rangle$ denotes a series of stationary atomic states with energies
$\omega_{n}$ and $\sigma_{nn}(\tau)=|n\rangle\langle n|$
 ($\hbar=c=1$). The free Hamiltonian of the quantum  field is given
 by
 \beq
 H_F (\tau) = \sum_k \omega_\k \,a^\dagger_\k a_\k {dt\/d
\tau} \;, \label{HF}
 \eeq
where  $\k$ denotes the wave vector and polarization of the field
modes.  The atom-field interaction Hamiltonian $H_I$ is given by
(\ref{HI}), which, in the frame of the atom, reduces to
 \beq
 H_I (\tau) =
-er_i(\tau)\,{\partial_i\phi(x(\tau))}=
-e\sum_{mn}(r_i)_{mn}\,\sigma_{mn}(\tau)\,{\partial_i\phi(x(\tau))}
\label{HI3} \;.
 \eeq
In what follows, we choose to work in this reference frame. The
Heisenberg equations of motion for the dynamical variables of the
 atom and the scalar field can be derived from the Hamiltonian
$H=H_{A}+H_{F}+H_{I}$: \beq {{d}\/{d\tau}}\sigma_{mn}(\tau) =
i(\omega_{m}-\omega_{n})\sigma_{mn}(\tau)-ie{\partial_i
\phi(x(\tau))}\bigg
[r_i(\tau),\sigma_{mn}(\tau)\bigg]\;,\label{eq4}
 \eeq
 \beq
{{d}\/{dt}}a_{\k}(t(\tau))=-i\omega_{\k}a_{\k}(t(\tau))-ier_i\bigg[
{\partial_i
\phi(x(\tau))},a_{\k}(t(\tau))\bigg]{{d\tau}\/{dt}}\;.\label{eq5}
 \eeq
In the solutions of the equations of motion, we can separate the
"free" and "source" parts,
 \beq \sigma_{mn}(\tau)
=\sigma_{mn}^{f}(\tau)+\sigma_{mn}^{s}(\tau), \qquad
a_{\k}(t(\tau))  = a_{\k}^{f}(t(\tau))+a_{\k}^{s}(t(\tau))\; ,
\label{eq6} \eeq
 where
 \beq
 \sigma_{mn}^{f}(\tau) =
\sigma_{mn}^f(\tau_{0})e^{i(\omega_{m}-\omega_{n})(\tau-\tau_{0})},
\qquad \sigma_{mn}^{s}(\tau)=-ie\int_{\tau_{0}}^{\tau}
d\tau'{\partial_j\phi^f(x(\tau'))}\bigg[r_j^f(\tau'),\sigma^f_{mn}(\tau)\bigg]\;,
\label{eq7}
 \eeq
 \beq
 a_{\k}^{f}(\tau) =
a_{\k}^f(t(\tau_{0}))e^{-i\omega_{\k}[t(\tau)-t(\tau_{0})]}\;,
\qquad a_{\k}^{s}(t(\tau))=-ie\int_{\tau_{0}}^{\tau}
d\tau'r_j^f(\tau')\bigg[{\partial_j\phi^f(x(\tau'))},a_{\k}^f(t(\tau))\bigg]\;.
\label{eq8}
 \eeq

\section{The contributions of vacuum fluctuation and radiation reaction to the rate of variation of the atomic energy}

 We assume that the initial state of the field is the vacuum $|0\rangle$, while the
atom is in the state $|b\rangle$. The equation of motion in the
interaction representation for an arbitrary atomic observable
$G(\tau)$, using symmetric ordering \cite{Dalibard82}, can be
split into the vacuum fluctuations and the radiation reaction
contributions,
 \beq
 {{d}\/{d\tau}}G(\tau)=
\bigg({{d}\/{d\tau}}G(\tau)\bigg)_{VF}+\bigg({{d}\/{d\tau}}G(\tau)
\bigg)_{RR}\;,
 \eeq
where
 \beq
 \bigg({{d}\/{d\tau}}G(\tau)\bigg)_{VF}=
-{{ie}\/{2}}\bigg\{{\partial_i
\phi^f(x(\tau))}[r_i(\tau),G(\tau)]+[r_i(\tau),G(\tau)]{\partial_i
\phi^f(x(\tau))}\bigg\}\;,\label{eq10}
 \eeq
  representing the contribution of the vacuum fluctuations and
 \beq
\bigg\{{{d}\/{d\tau}}G(\tau)\bigg\}_{RR}=
-{{ie}\/{2}}\bigg\{{\partial_i
\phi^s(x(\tau))}[r_i(\tau),G(\tau)]+[r_i(\tau),G(\tau)]{\partial_i
\phi^s(x(\tau))}\bigg\},\label{eq11}
 \eeq
 denoting that of the radiation reaction.

Our main aim now is to identify the contributions of vacuum
fluctuations and radiation reaction in the evolution of the atom's
excitation energy, which is given by the expectation value of
$H_{A}$. Separating $r_{i}(\tau)$ and $\sigma_{nn}(\tau)$ into
their free part and source part and taking the vacuum expectation
value, we can obtain, in a perturbation treatment up to order
$e^2$,
 \beq
 \bigg \langle0\bigg |{{dH_{A}(\tau)}\/{d\tau}}\bigg
|0\bigg \rangle_{VF}=-e^2\int_{\tau_{0}}^\tau d\tau'
C_{ij}^F(x(\tau),x(\tau'))\bigg[r_{j}^f(\tau'),[r_i^f(\tau),
\sum_n\omega_{n}\sigma_{nn}^f(\tau)]\bigg]\;,\label{eq12}
 \eeq
 \beq
 \bigg \langle0\bigg |{{dH_{A}(\tau)}\/{d\tau}}\bigg |0\bigg
\rangle_{RR}=e^2\int_{\tau_{0}}^\tau d\tau'
\chi_{ij}^F(x(\tau),x(\tau'))\bigg\{r_{j}^f(\tau'),[r_i^f(\tau),
\sum_n\omega_{n}\sigma_{nn}^f(\tau)]\bigg\}\;.\label{eq13}
 \eeq
The statistical functions, $C_{ij}^{F}(x(\tau),x(\tau'))$ and
$\chi_{ij}^{F}(x(\tau),x(\tau'))$,  of the field derivative are
written as
 \beq
C_{ij}^{F}(x(\tau),x(\tau'))={{1}\/{2}}\bigg\langle0\bigg|\bigg\{{\partial_i
\phi^f(x(\tau))},{\partial_j
\phi^f(x(\tau'))}\bigg\}\bigg|0\bigg\rangle\;,\label{eq14}
 \eeq
 \beq
 \chi_{ij}^{F}(x(\tau),x(\tau'))={{1}\/{2}}\bigg\langle
0\bigg|\bigg[{\partial_i\phi^f(x(\tau))},{\partial_j\phi^f(x(\tau'))
}\bigg]\bigg|0\bigg\rangle\;.\label{eq15}
 \eeq
Since we are interested in the evolution of expectation values of
atomic observables,  we take the expectation value of
Eqs.(\ref{eq12}) and (\ref{eq13}) in the atom's state $|b\rangle$.
Using the Heisenberg equation of motion, we can replace the
commutator $[r_{i}^f(\tau),
\sum_{n}\omega_{n}\sigma_{nn}^f(\tau)]$ by
$i{{d}\/{d\tau}}r_{i}^{f}$, and obtain
 \beq
 \bigg\langle{{dH_{A}(\tau)}\/{d\tau}}\bigg \rangle_{VF}=2ie^2
\int_{\tau_{0}}^{\tau}d\tau'C_{ij}^{F}(x(\tau),x(\tau'))
{{d}\/{d\tau}}(\chi_{ij}^{A})_b(\tau,\tau')\;,\label{eq16}
 \eeq
 \beq
\bigg \langle{{dH_{A}(\tau)}\/{d\tau}}\bigg \rangle_{RR}=2ie^2
\int_{\tau_{0}}^{\tau}d\tau'\chi_{ij}^{F}(x(\tau),x(\tau'))
{{d}\/{d\tau}}(C_{ij}^{A})_b(\tau,\tau')\;,\label{eq17}
 \eeq
 where $|\rangle=|b,0\rangle$. Here the symmetric correlation function and
linear susceptibility of the atom are given analogously to
Eqs.(\ref{eq14}) and (\ref{eq15}) by
 \beq
(C_{ij}^{A})_b(\tau,\tau')={{1}\/{2}}\langle b| \{r_{i}^{f}
(\tau),r_{j}^f(\tau')\}|b\rangle\;,\label{eq18}
 \eeq
 \beq (\chi_{ij}^{A})_b(\tau,\tau')={{1}\/{2}}\langle b|[r_{i}^{f}
(\tau),r_{j}^f(\tau')]|b\rangle\;.\label{eq19}
 \eeq
These functions do not depend on the trajectory of the atom and
they only characterize  the atom itself. Now we give the explicit
forms of the statistical function of the atom
 \beq
(C_{ij}^{A})_b(\tau,\tau')={{1}\/{2}}\sum_{d}\bigg[\langle
b|r_i^f(\tau_0)|d\rangle \langle d|r_j^f(\tau_0)|b\rangle
e^{i\omega_{bd}(\tau-\tau')}+\langle b|r_j^f(\tau_0)|d\rangle
\langle d|r_i^f(\tau_0)|b\rangle
e^{-i\omega_{bd}(\tau-\tau')}\bigg]\;,\label{eq20}
 \eeq
 \beq
 (\chi_{ij}^{A})_b(\tau,\tau')={{1}\/{2}}\sum_{d}\bigg[\langle
b|r_i^f(\tau_0)|d\rangle \langle d|r_j^f(\tau_0)|b\rangle
e^{i\omega_{bd}(\tau-\tau')}-\langle b|r_j^f(\tau_0)|d\rangle
\langle d|r_i^f(\tau_0)|b\rangle
e^{-i\omega_{bd}(\tau-\tau')}\bigg]\;,\label{eq21} \eeq
 where $\omega_{bd}=\omega_b-\omega_d$ and the sum extends over a
complete set of atomic states.

\section{Spontaneous emission from a uniformly moving atom}

To begin with, we first apply the above developed formalism to
consider the spontaneous emission from an inertial atom moving in
the $x$-direction with a constant velocity $v$. The atom's
trajectory is given by
 \beq
  t(\tau)= \gamma\tau, \qquad x(\tau) =
x_0+v\gamma\tau, \qquad y(\tau) = z(\tau)=0\;, \label{eq22}
 \eeq
where $\gamma=(1-v^2)^{{-1}\/{2}}$. The correlation function of
the field derivative in the frame of the atom can be readily
calculated as follows
 \bea
 \bigg \langle0\bigg
|{\partial_i\phi^f(x(\tau))}{\partial_j\phi^f(x(\tau'))}\bigg
|0\bigg \rangle
={{\delta_{ij}}\/{2\pi^2(\tau-\tau'-i\epsilon)^4}}\;.\label{eq23}
 \eea
From Eq.~(\ref{eq23}), we get the symmetric function
 \beq
C_{ij}^F(x(\tau),x(\tau'))={{\delta_{ij}}\/{4\pi^2}}\bigg[{{1}
\/({\tau-\tau'-i\epsilon})^4}+
{{1}\/({\tau-\tau'+i\epsilon})^4}\bigg]\;, \label{eq24}
 \eeq
and the linear susceptibility
 \beq
 \chi^F_{ij}(x(\tau),x(\tau'))=-{{i}\/{12\pi}}\delta_{ij}
\delta^{'''}(\tau-\tau')\;, \label{eq25}
 \eeq
where $\delta^{'''}(\tau-\tau')$ is the third derivative of the
Dirac delta function. With a substitution $u=\tau-\tau'$, we
obtain the contribution of the vacuum fluctuations to the rate of
change of the atomic energy with Eq.~(\ref{eq16})
 \bea
 \bigg \langle
{{dH_A(\tau)}\/{d\tau}}\bigg
\rangle_{VF}&=&-{{e^2}\/{4\pi^2}}\sum_d |\langle
b|{\mathbf{r}}^f(\tau_0)|d\rangle|^2\omega_{bd}\int_{-\infty}^{+\infty}du\,
e^{i\omega_{bd}u}\bigg[{{1}\/{(u-i\epsilon)^4}}+{{1}\/{(u+i\epsilon)^4}}\bigg]
\nonumber\\&=&-{{e^2}\/{12\pi}}\sum_{\omega_{bd}>0}|\langle
b|{\mathbf{r}}^f(\tau_0)|d\rangle|^2\omega_{bd}^4+{{e^2}\/{12\pi}}
\sum_{\omega_{bd}<0}|\langle
b|{\mathbf{r}}^f(\tau_0)|d\rangle|^2\omega_{bd}^4\;,\label{eq26}
 \eea
 and that of the radiation reaction
 \bea
 \bigg \langle {{dH_A(\tau)}\/{d\tau}}\bigg
\rangle_{RR}&=&{{ie^2}\/{12\pi^2}}\sum_d |\langle
b|{\mathbf{r}}^f(\tau_0)|d\rangle|^2\omega_{bd}\int_{-\infty}^{+\infty}du\,
e^{i\omega_{bd}u}\delta^{'''}(u)
\nonumber\\&=&-{{e^2}\/{12\pi}}\biggl(\sum_{\omega_{bd}>0}|\langle
b|{\mathbf{r}}^f(\tau_0)|d\rangle|^2\omega_{bd}^4+\sum_{\omega_{bd}<0}|\langle
b|{\mathbf{r}}^f(\tau_0)|d\rangle|^2\omega_{bd}^4\bigg)\;,\label{eq27}
 \eea
where we have extended the range of integration to infinity for
sufficiently long times. After the calculation of the integrals,
we obtain
 \bea
 \bigg \langle {{dH_A(\tau)}\/{d\tau}}\bigg
\rangle_{tot}=-{{e^2}\/{6\pi}}\sum_{\omega_{bd}>0}|\langle
b|{\mathbf{r}}^f(\tau_0)|d\rangle|^2\omega_{bd}^4\;.\label{eq28}
 \eea
This result, except for a numerical factor, is the same as that
obtained in Ref.~\cite{Audretsch94} where a monopole like
interaction with the scalar field is assumed. This shows that, in
terms of the spontaneous emission rate, an inertial atom does not
tell the difference of the type of interaction between the atom
and the quantum field, be it of the linear monopole-field or
dipole-derivative type. However, things will be different when we
come to the case of an accelerated atom, as we will demonstrate
next

\section{Spontaneous excitation from a uniformly accelerated atom}

Let us now turn our attention to the case in which the atom is
uniformly accelerated in the $x$-direction. The atom's trajectory
is now given by
 \beq
 t(\tau)= {1\/a} \sinh a \tau, \qquad x(\tau) =
{1\/a} \cosh a \tau, \qquad y(\tau) = z(\tau)=0\;.\label{eq29}
 \eeq
The two point function of the field derivatives on the atom's
trajectory can be evaluated in the frame of the atom to get
 \beq
 \bigg \langle0\bigg |{\partial_x
\phi^f(x(\tau))}{\partial_x \phi^f(x(\tau'))}\bigg |0\bigg
\rangle={{a^4}\/{32\pi^2\sinh^4({{a(\tau-\tau')}\/{2}}-i\epsilon)}}
\bigg\{1-2\sinh^2{{a(\tau-\tau')}\/{2}}\bigg\}\;,\label{eq30}
 \eeq
\bea \bigg \langle0\bigg |{\partial_y \phi^f(x(\tau))}{\partial_y
\phi^f(x(\tau'))}\bigg |0\bigg \rangle&=&\bigg \langle0\bigg
|{\partial_z \phi^f(x(\tau))}{\partial_z \phi^f(x(\tau'))}\bigg
|0\bigg
\rangle\nonumber\\&=&{{a^4}\/{32\pi^2\sinh^4({{a(\tau-\tau')}\/{2}}-i\epsilon)}}
\;,\label{eq31}
 \eea
 and
 \beq
 \bigg \langle0\bigg |{\partial_i
\phi^f(x(\tau))}{\partial_j \phi^f(x(\tau'))}\bigg |0\bigg
\rangle_{i\neq j}=0\;.\label{eq32}
 \eeq
Consequently, we obtain the symmetric correlation function
 \beq
 C_{xx}^{F}(x(\tau),x(\tau'))={{a^4[1-2\sinh^2{{a(\tau-\tau')}
\/{2}}]}\/{64\pi^2}}
\bigg[{{1}\/{\sinh^4({{a(\tau-\tau')}\/{2}}-i\epsilon)}}
+{{1}\/{\sinh^4({{a(\tau-\tau')}\/{2}}+i\epsilon)}}\bigg]
\;,\label{eq33} \eeq
  \beq
C_{yy}^{F}(x(\tau),x(\tau'))=C_{zz}^{F}(x(\tau),x(\tau'))={{a^4}\/{64\pi^2}}
\bigg[{{1}\/{\sinh^4({{a(\tau-\tau')}\/{2}}-i\epsilon)}}
+{{1}\/{\sinh^4({{a(\tau-\tau')}\/{2}}+i\epsilon)}}\bigg]
\;,\label{eq34}
 \eeq
 and the linear susceptibility
\beq
\chi_{xx}^{F}(x(\tau),x(\tau'))=-{{i(1-2\sinh^2{{a(\tau-\tau')}\/{2}})}\/
{2\pi\cosh{{a(\tau-\tau')}\/{2}}(5+\cosh^2{{a(\tau-\tau')}\/{2}})}}
\delta^{'''}(\tau-\tau')\;,\label{eq35} \eeq and
 \beq
\chi_{yy}^{F}(x(\tau),x(\tau'))=\chi_{zz}^{F}(x(\tau),x(\tau'))=
-{{i}\/
{2\pi\cosh{{a(\tau-\tau')}\/{2}}(5+\cosh^2{{a(\tau-\tau')}\/{2}})}}
\delta^{'''}(\tau-\tau')\;.\label{eq36}
 \eeq
 Since the polarization direction of the atom can be
arbitrary, in general, the polarization can have non-zero
components in both the direction of acceleration and that which is
perpendicular to it. So calculations have to be carried out for
all non-zero field derivative statistical functions. Then, it is
easy to show the contribution of the vacuum fluctuations
associated with the $xx$ component of the statistical functions is
given by
 \bea
 \bigg \langle {{dH_A(\tau)}\/{d\tau}}\bigg
\rangle_{VFxx}&=&-{{e^2a^4}\/{64\pi^2}}\sum_d\omega_{bd} |\langle
b|r_x^f(\tau_0)|d\rangle|^2\int_{-\infty}^{+\infty}du\,
e^{i\omega_{bd}u}
(1-2\sinh^2{{au}\/{2}})\nonumber\\&\bigg[&{{1}\/{\sinh^4({{au}\/{2}}-i\epsilon)}}
+{{1}\/{\sinh^4({{au}\/{2}}+i\epsilon)}}\bigg]\; ,\label{eq37}
 \eea
and that with other non-zero components are
 \bea \bigg \langle {{dH_A(\tau)}\/{d\tau}}\bigg
\rangle_{VFii}&=&-{{e^2a^4}\/{64\pi^2}}\sum_d\omega_{bd} |\langle
b|r_i^f(\tau_0)|d\rangle|^2\int_{-\infty}^{+\infty}du\,
e^{i\omega_{bd}u}
\nonumber\\&\bigg[&{{1}\/{\sinh^4({{au}\/{2}}-i\epsilon)}}
+{{1}\/{\sinh^4({{au}\/{2}}+i\epsilon)}}\bigg] ,\label{eq38}
 \eea
for $i\neq x$. With the help of the follow integrals which can be
easily calculated using residue theorem \bea
 &&\int_{-\infty}^{+\infty}du\;
e^{i\omega_{bd}u}
(1-2\sinh^2{{au}\/{2}})\bigg[{{1}\/{\sinh^4({{au}\/{2}}-i\epsilon)}}
+{{1}\/{\sinh^4({{au}\/{2}}+i\epsilon)}}\bigg] \nonumber\\&=&
{{16\pi}\/{3a^4}}|\omega_{bd}|
(\omega_{bd}^2+4a^2)\,{{e^{(2\pi|\omega_{bd}|/a)}+1}\/
{e^{(2\pi|\omega_{bd}|/a)}-1}}\;,\label{eq39}
 \eea
 \bea
 \int_{-\infty}^{+\infty}du\; e^{i\omega_{bd}u}
\bigg[{{1}\/{\sinh^4({{au}\/{2}}-i\epsilon)}}
+{{1}\/{\sinh^4({{au}\/{2}}+i\epsilon)}}\bigg] =
{{16\pi}\/{3a^4}}|\omega_{bd}|
(\omega_{bd}^2+a^2)\,{{e^{(2\pi|\omega_{bd}|/a)}+1}\/
{e^{(2\pi|\omega_{bd}|/a)}-1}}\;,\label{eq40}
 \eea
 we find
 \bea
 \bigg \langle {{dH_A(\tau)}\/{d\tau}}\bigg
\rangle_{VFxx}=&-&{{e^2}\/{12\pi}}\sum_{\omega_{bd}>0} |\langle b|
r_x^f(\tau_0)|d\rangle|^2\omega_{bd}^4\,\biggl(1+{{4a^2}\/{\omega_{bd}^2}}\biggr)\bigg[1+
{{2}\/{e^{(2\pi\omega_{bd}/a)}-1}}\bigg]\nonumber\\&+&{{e^2}\/{12\pi}}
\sum_{\omega_{bd}<0}|\langle b|
r_x^f(\tau_0)|d\rangle|^2\omega_{bd}^4\,\biggl(1+{{4a^2}\/{\omega_{bd}^2}}\biggr)\bigg[1+
{{2}\/{e^{(2\pi\omega_{bd}/a)}-1}}\bigg]\; ,\label{eq41}
 \eea
and
 \bea
 \bigg \langle {{dH_A(\tau)}\/{d\tau}}\bigg
\rangle_{VFii}=&-&{{e^2}\/{12\pi}}\sum_{\omega_{bd}>0} |\langle b|
r_i^f(\tau_0)|d\rangle|^2\omega_{bd}^4\,\biggl(1+{{a^2}\/{\omega_{bd}^2}}\biggr)\bigg[1+
{{2}\/{e^{(2\pi\omega_{bd}/a)}-1}}\bigg]\nonumber\\&+&{{e^2}\/{12\pi}}
\sum_{\omega_{bd}<0}|\langle b|
r_i^f(\tau_0)|d\rangle|^2\omega_{bd}^4\,\biggl(1+{{a^2}\/{\omega_{bd}^2}}\biggr)\bigg[1+
{{2}\/{e^{(2\pi\omega_{bd}/a)}-1}}\bigg] ,\label{eq42}
 \eea
 for $i\neq x$.

 Vacuum fluctuations  tend to excite an accelerated ground-state atom and de-excite it
 in the excited state and the probability of these processes are enhanced by the acceleration dependent correction
 terms as compared to the inertial case.  Notice the appearance of nonthermal term
 proportional to $a^2$ as compared with the purely thermal result obtained in Ref. \cite{Audretsch94} for the monopole-field interaction.
An interesting feature worth being noted is that the nonthermal
correction for a longitudinally polarized atom (polarized in the
direction of acceleration) is four time that for a transversely
polarized atom (polarized in the perpendicular direction).

Similarly, we can find the contributions of the radiation reaction
as follows
 \bea
 \bigg \langle {{dH_A(\tau)}\/{d\tau}}\bigg
\rangle_{RRxx}&=&{{ie^2}\/{2\pi}}\sum_d|\langle
b|r_x^f(\tau_0)|d\rangle|^2\omega_{bd}\int_{-\infty}^{+\infty}du\,\delta^{'''}(u)\,
{{e^{i\omega_{bd}u}(1-2\sinh^2{{au}\/{2}})}\/{\cosh{{au}\/{2}}(5+\cosh^2{{au}\/{2}})}}
\nonumber\\&=&-{{e^2}\/{12\pi}}\biggl\{\;\sum_{\omega_{bd}>0}|\langle
b|r_x^f(\tau_0)|d\rangle|^2\omega_{bd}^4\,\biggl(1+{{4a^2}\/{\omega_{bd}^2}}\biggr)
\nonumber\\
&&+\sum_{\omega_{bd}<0}|\langle
b|r_x^f(\tau_0)|d\rangle|^2\omega_{bd}^4\,\biggl(1+{{4a^2}\/{\omega_{bd}^2}}\biggr)\biggr\}\;,\label{eq44}
 \eea
and
 \bea
 \bigg \langle {{dH_A(\tau)}\/{d\tau}}\bigg
\rangle_{RRii}&=&{{ie^2}\/{2\pi}}\sum_d|\langle
b|r_i^f(\tau_0)|d\rangle|^2\omega_{bd}\int_{-\infty}^{+\infty}du\,\delta^{'''}(u)\,
{{e^{i\omega_{bd}u}}\/{\cosh{{au}\/{2}}(5+\cosh^2{{au}\/{2}})}}
\nonumber\\&=&-{{e^2}\/{12\pi}}\,\biggl\{\,\sum_{\omega_{bd}>0}|\langle
b|r_i^f(\tau_0)|d\rangle|^2\omega_{bd}^4\,\biggl(1+{{a^2}\/{\omega_{bd}^2}}\biggr)
\nonumber\\
&&+\biggl\{\,\sum_{\omega_{bd}<0}|\langle
b|r_i^f(\tau_0)|d\rangle|^2\omega_{bd}^4\,\biggl(1+{{a^2}\/{\omega_{bd}^2}}\biggr)
\biggr\}\;,\label{eq45}
 \eea
for $i\neq x$.
 The above result shows that the contribution of radiation reaction to the rate
 of change of the atomic energy always leads to a loss of energy
 of atoms and it is affected by the acceleration. This is to be
 contrasted to the case with a monopole like interaction between the atom and the
 field where it has been demonstrated that first for uniformly accelerated atoms \cite{Audretsch94}
 and then for accelerated atoms on arbitrary stationary
trajectory \cite{Audretsch95B}, the contribution of radiation
reaction is generally not altered from its inertial value. This
along with what we have found elsewhere when interaction with
electromagnetic fields is considered \cite{Yu06} suggests that the
non-alteration of the contributions of radiation reaction to rate
of change of the mean atomic energy from its inertial value is a
property which is only unique to the case considered in
Ref.~\cite{Audretsch94}.  Again, the acceleration-induced
correction for the longitudinally polarized atoms is four times
that for transversely polarized ones.

Finally, we add up the contributions of vacuum fluctuations and
radiation reaction to obtain the total rate of change of the
atomic excitation energy:
 \bea
 \bigg \langle {{dH_A(\tau)}\/{d\tau}}\bigg
\rangle_{tot}=&-&{{e^2}\/{6\pi}}
\sum_{\omega_{bd}>0}\sum_{i}|\langle
b|r_i^f|d\rangle|^2\omega_{bd}^4\, f_i(a,\omega_{bd}
)\,\bigg(1+{{1}\/{e^{(2\pi\omega_{bd}/a)}-1}}\bigg)
\nonumber\\&+&{{e^2}\/{6\pi}} \sum_{\omega_{bd}<0}\sum_{i}|\langle
b|r_i^f|d\rangle|^2\omega_{bd}^4\,f_i(a,\omega_{bd})\,{{1}\/{e^{(2\pi|\omega_{bd}|/a)}-1}}
,\label{eq47} \eea
 where functions $f_i(a,\omega_{bd})$ are defined as follows
 \beq
 f_x(a,\omega_{bd})= 1+{4a^2\/\omega_{bd}^2}\;, \quad\quad\ f_y(a,\omega_{bd})=
 f_z(a,\omega_{bd})=1+{a^2\/\omega_{bd}^2}\;.
 \eeq
For a ground-state atom, although both contributions of the vacuum
fluctuations and radiation are altered for accelerated atoms with
the dipole-derivative coupling, as opposed to no change in the
contribution of radiation reaction in the monopole-field coupling
\cite{Audretsch94}, they conspire to change in such a way that the
delicate balance between the vacuum fluctuations and radiation
reaction no longer exists. There is a positive contribution from
the $\omega_{b}< \omega_{d}$ term, therefore transitions of
ground-state atoms to excited states are allowed to occur even in
vacuum.

\section{Conclusions}

In conclusion, assuming a linear coupling between the dipole
moment of a multi-level atom and the derivative of a  massless
quantum scalar field, we have calculated the contributions of
vacuum fluctuations and radiation reaction to the rate of change
of the atomic energy of a uniformly accelerated atom. In contrast
to the case where a monopole like interaction is assumed
\cite{Audretsch94}, there appear extra corrections proportional to
$a^2$, in addition to corrections which can be viewed as a result
of an ambient thermal bath at the Unruh temperature $T=a/2\pi$ as
compared with the inertial case.  Let us note that similar terms
have also been found in the response function for a free falling
Unruh detector in de Sitter space interacting with minimally
coupled massless scalar fields~\cite{GP04}. The deviation from
pure thermal behavior of the spontaneous excitation rate of the
uniformly accelerated atom in the our case and the response
functions of a free falling Unruh detector in de Sitter space by
no means imply the exact final thermal equilibrium is not
achieved~\cite{GP04,Brout}. As a matter of fact, with the
transition probabilities for the uniformly accelerated atom which
can be found from our calculation in the preceding Section, one
can show, by the same argument as that in Ref~\cite{GP04},  that
exact thermal equilibrium will be established at the Unruh
temperature. However, although a discrepancy of the excitation
rate in the present case with the pure thermal one neither leads
to thermal non-equilibrium nor a different thermal equilibrium
temperature, but, with different behaviors of the transition
probabilities of the atoms, it does imply a clear difference in
how atomic transitions occur and how the equilibrium is reached.

It is interesting to note that even if the atom is isotropically
polarized, the acceleration induced correction terms in addition
to the pure thermal one show anisotropy with the contribution
arising from longitudinal polarization being four times that from
the transverse polarization. This can probably understood as a
result of the fact that there is one distinguished spatial
direction, namely, the direction of acceleration.  This is to be
contrasted with the case in which dipole coupling with
electromagnetic field is considered \cite{Yu06} and there one
finds that the acceleration induced terms are isotropic. However,
it should be pointed out that the isotropy associated with
electromagnetic field is probably a property unique to the four
dimensions. An example of this is that the electromagnetic vacuum
noise is isotropic in four dimensions but not in higher dimensions
\cite{Takagi86}. Finally, our results show that the effect of
acceleration on the rate of change of the mean atomic energy is
dependent not only  on the quantum field to which the atom is
coupled (electromagnetic vs scalar), but also on the type of the
interaction ( monopole-field vs dipole-derivative ) even if the
same scalar quantum field is considered.

\begin{acknowledgments}
This work was supported in part  by the National Natural Science
Foundation of China  under Grants No.10375023 and No.10575035, and
the Program for NCET (No. 04-0784).
\end{acknowledgments}


\end{document}